\documentclass[twocolumn]{article}
\usepackage{graphicx}
\usepackage{url}
\usepackage{array}
\usepackage{pbox}
\usepackage{tikz}
\usepackage{tikz-qtree}
\usepackage{multicol}

\usetikzlibrary{arrows}

\tikzstyle{tpq}=[level distance=17, every node/.style={inner sep=0,outer sep=1}, sibling distance=0.4cm, edge from parent/.style={draw, edge from parent path={(\tikzparentnode)--(\tikzchildnode)}}]

\tikzset{axad/.append style={double, double distance=0.5mm}}

\newcommand{\qnodei}[1]{{\small\texttt{\#}\ensuremath{\mathsf{#1}}}}

\newcommand*\circled[1]{\tikz[baseline=(char.base)]{
		\node[shape=circle,draw,inner sep=0.1pt] (char) {#1};}}

\newcommand{\qnodeo}[1]{\circled{{\small\texttt{\#}\ensuremath{\mathsf{#1}}}}}

\newcommand{\clause}[1]{`#1'}

\newcommand{\algo}[1]{\emph{#1}}

\begin{document}

\title{RadegastXDB -- Prototype of Native XML Database Management System: Technical Report}

\author{
	\begin{tabular}{p{5cm}p{5cm}p{5cm}}
		\centering
		\noindent Petr Luk\'{a}\v{s} \linebreak
		\noindent \texttt{petr.lukas@vsb.cz}
		&
		\centering
		\noindent Radim Ba\v{c}a \linebreak
		\noindent \texttt{radim.baca@vsb.cz}
		&
		\centering
		\noindent Michal Kr\'{a}tk\'{y} \linebreak
		\noindent \texttt{michal.kratky@vsb.cz}
	\end{tabular} \\ \\
	Department of Computer Science \\
	Faculty of Electrical Engineering and Computer Science \\
	VSB -- Technical University of Ostrava
}

\twocolumn[
\maketitle
]

\begin{abstract}
A lot of advances in the processing of XML data have been proposed in last two decades. There were many approaches focused on the efficient processing of twig pattern queries (TPQ). However, including the TPQ into an XQuery compiler is not a straightforward task and current XML DBMSs process XQueries without any TPQ detection. In this paper, we demonstrate our prototype of a~native XML DBMS called RadegastXDB that uses a~TPQ detection to accelerate structural XQueries. Such a~detection allows us to utilize state-of-the-art TPQ processing algorithms. Our experiments show that, for the structural queries, these algorithms and state-of-the-art XML indexing techniques make our prototype faster than all of the current XML DBMSs, especially for large data collections. We also show that using the same techniques is also efficient for the processing of queries with value predicates.
\end{abstract}

\section{Introduction}
\label{sec:introduction}

A~lot of advances in the processing of XML data have been proposed in last two decades. Especially in 2000 -- 2010, there were many approaches focused on an efficient processing of XQueries modeled by \emph{twig pattern queries} (TPQ) (e.g., \cite{ZNDLL01,KJK02,BSK02,wpj03,So+06,Re06,Qin+07,WH09}). In general, there are two major groups of TPQ processing algorithms: binary structural joins \cite{KJK02,al2002multi,wpj03,GKT03} and holistic twig joins \cite{BSK02,So+06,Qin+07,Ba+13}, where the latter group is considered as the state-of-the-art. However, the most of the current XML database management systems (DBMSs) do not utilize holistic twig joins, since these DBMSs are not capable to detect TPQs in XQueries. Instead, they rely on rather naive techniques such as nested loops or the traditional relational merge join algorithms. In other words, they ignore the most of the advances in the XML query processing introduced in last two decades and, therefore, they perform poorly even on simple structural queries on large data collections.

\begin{figure}[htb]
	\centering
	\begin{tabular}{
			>{\centering\arraybackslash} m{4cm}
			>{\centering\arraybackslash} m{3cm}}
		\textsf{\pbox{6cm}{
				\textbf{for} \$i \textbf{in} //r/a \newline
				\textbf{for} \$j \textbf{in} \$i/b//c \newline
				\textbf{for} \$k \textbf{in} \$i//d[./e \textbf{and} .//f] \newline 
				\textbf{return} (\$i, \$j, \$k)
			}}
			&
			\begin{tikzpicture}[tpq]		
			\Tree
			[.\qnodei{r}
			[.\node(a){\qnodeo{a}};
			[.\node(b){\qnodei{b}};
			\edge[axad];
			[.\node(c){\qnodeo{c}};
			]
			]
			\edge[axad];					
			[.\node(d){\qnodeo{d}};
			[.\qnodei{e}
			]
			\edge[axad];
			[.\qnodei{f}
			]
			]				
			]
			]			
			\end{tikzpicture} \\
			(a) Sample XQuery & (b) Corresponding TPQ
		\end{tabular}
		
		\caption{Sample XQuery and its TPQ}
		\label{fig:sample_tpq}
	\end{figure}
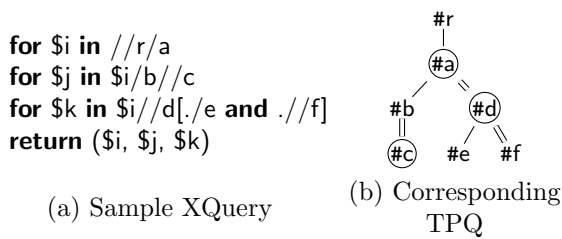

A~TPQ is a~rooted labeled tree, where each node corresponds to one location step in an XQuery. A~sample TPQ is illustrated in Figure~\ref{fig:sample_tpq}b and it corresponds to the XQuery in Figure~\ref{fig:sample_tpq}a. The single and double lined edges represent the parent-child (PC) and ancestor-descendant (AD) structural relationships corresponding to the \texttt{child} and \texttt{descendant} axes, respectively\footnote{For the sake of simplicity, we consider only these two XPath axes as far as it is common for the most of the XML query processing approaches.}. We call \emph{query nodes} the nodes in a~TPQ and we denote them by the `\#' character. Additionally, the circled query nodes represent \emph{output query nodes} (also called \emph{extraction points} \cite{mathis2006hash}) which correspond to the last location steps in the \clause{for} clauses. In a~nutshell, the processing of a~TPQ means to find all mappings from the TPQ to an XML document such that the query nodes are mapped to XML nodes of the corresponding name and these XML nodes satisfy the relationships specified by the query edges. For more details about the processing of a~TPQ, we refer to~\cite{bavca2017structural}.

	In this paper, we demonstrate our prototype of a native XML DBMS called RadegastXDB. The prototype mainly focuses on an efficient processing of structural XQueries which is primarily possible due to the proper detection of TPQs. Such a~detection allows us to utilize state-of-the-art query processing algorithms. Our experiments show that, for the structural queries,  these algorithms and state-of-the-art XML indexing techniques make our prototype faster than all of the current XML DBMSs. We also show that using the same techniques is also efficient for the processing of queries with value predicates.
	
	The rest of this paper is organized as follows. In Section~\ref{sec:architecture}, we outline the architecture of our prototype including index data structures and query processing techniques, and in Section~\ref{sec:experiments}, we provide an experimental comparison with the current XML DBMSs. Section~\ref{sec:conclusions} concludes the paper.
	
\section{Architecture}
\label{sec:architecture}

The RadegastXDB native XML DBMS consists of two main subsystems: the storage and the XQuery processor. The storage is a~set of indexes to store a~collection of XML documents and to perform low-level data access operations. The XQuery processor represents the front-end of our prototype. We outline the storage and the XQuery processor in the following subsections. 


%
%
%
%
%
%

\subsection{Storage}

The storage consists of three indexes, namely: the document index, the partition index, and the value index. Let us note that similar indexes are described in \cite{WH10}. 

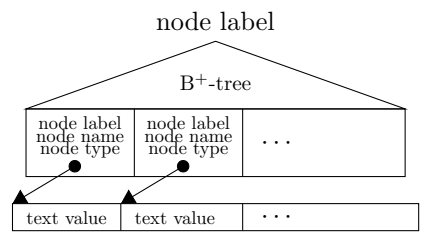
\begin{figure}[htb]
	\centering
	\begin{tikzpicture}[scale=0.9]
	\draw (4.6,1.4) -- (1.8,0.4) -- (7.4,0.4) -- (4.6,1.4);
	\node [scale = 0.8] at (4.6,0.8) {B$^+$-tree};
	\node at (4.6,1.7) {node label};
	\draw  (1.8,0.4) rectangle (7.4,-0.6);
	
	\node [scale=0.7] at (2.6,0.2) {node label};
	\node [scale=0.7] at (2.6,0) {node name};
	\node [scale=0.7] at (2.6,-0.2) {node type};
	\draw (3.4,0.4) -- (3.4,-0.6);
	
	\node [scale=0.7] at (4.2,0.2) {node label};
	\node [scale=0.7] at (4.2,0) {node name};
	\node [scale=0.7] at (4.2,-0.2) {node type};
	\draw (5,0.4) -- (5,-0.6);
	
	\node at (5.5,-0.1) {$\cdots$};

	\draw  (1.6,-1) rectangle (7.6,-1.4);
	\node [scale=0.7] at (2.4,-1.2) {text value};
	\node [scale=0.7] at (4,-1.2) {text value};
	\draw (3.2,-1) -- (3.2,-1.4);
	\draw (5,-1) -- (5,-1.4);
	double distance=2pt
	\node at (5.5,-1.2) {$\cdots$};
	
	\draw [*-triangle 60](2.6,-0.4) -- (1.6,-1);
	\draw [*-triangle 60](4.2,-0.4) -- (3.2,-1);
	\end{tikzpicture}
	\caption{Document index}
	\label{fig:document_index}
\end{figure}

The document index (see Figure~\ref{fig:document_index}) represents the main index that stores all data of any inserted XML document. It maps \emph{node labels} to the corresponding XML nodes. The node labels have two purposes: (1) to identify XML nodes uniquely and (2) to resolve structural relationships in a~constant time\footnote{We utilize the Containment labeling scheme \cite{ZNDLL01}.}. The document index consists of a~B$^+$-tree and a~paged array, where a~node label is the~key of the B$^+$-tree. Each leaf entry of the B$^+$-tree also includes an XML node name, a~type of the XML node (element, attribute, text, etc.), and a~pointer to the paged array where values of attributes and text nodes are stored. We can either perform a~point query on the B$^+$-tree to retrieve data of a~specific XML node or a~range query to retrieve data of all descendants of an XML node.

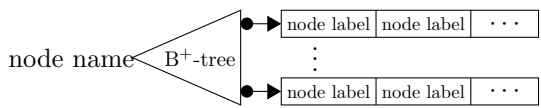
\begin{figure}[htb]
	\centering
	\begin{tikzpicture}[scale=0.9]	
	
	\draw (2.6,-3.3) -- (4.2,-2.6) -- (4.2,-4) -- (2.6,-3.3);
	\node [scale=0.8] at (3.6,-3.3) {B$^+$-tree};
	\node at (1.7,-3.3) {node name};
	
	\draw [*-triangle 60](4.2,-2.8) -- (4.8,-2.8);
	\draw  (4.8,-2.6) rectangle (8.6,-3);
	\node [scale=0.7] at (5.5,-2.8) {node label};
	\node [scale=0.7] at (6.9,-2.8) {node label};
	
	\draw (6.2,-2.6) -- (6.2,-3);
	\draw (7.6,-2.6) -- (7.6,-3);
	
	\node at (8.1,-2.8) {$\cdots$};
	
	\node at (5.3,-3.2) {$\vdots$};
	
	\draw [*-triangle 60](4.2,-3.8) -- (4.8,-3.8);
	\draw  (4.8,-3.6) rectangle (8.6,-4);
	\node [scale=0.7] at (5.5,-3.8) {node label};
	\node [scale=0.7] at (6.9,-3.8) {node label};
	
	\draw (6.2,-3.6) -- (6.2,-4);
	\draw (7.6,-3.6) -- (7.6,-4);
	
	\node at (8.1,-3.8) {$\cdots$};	
	
	\end{tikzpicture}
	\caption{Partition index}
	\label{fig:partition_index}
\end{figure}

The partition index (see Figure~\ref{fig:partition_index}) is also a~combination of a~B$^+$-tree and a~paged array, but an XML node name is the key of the B$^+$-tree and each leaf entry of the B$^+$-tree points to a~distinct paged array of node labels of the specific name. In such a~way, the partition index can retrieve a~list of node labels corresponding to a~specific name which mainly supports the processing of structural XQueries.

Finally, the value index has a~similar structure as the partition index, but the key of the B$^+$-tree is composed of an XML node name and value. Therefore, it supports the processing of XQueries with value predicates including both equality and inequality comparisons.



\subsection{XQuery Processor}
\label{sec:xquery_processor}

The XQuery processor is based on the Galax XQuery algebra \cite{Re06}. The prototype currently does not support the whole XQuery, since we mainly focus on the processing of structural queries. These queries can be modeled using TPQs which we detect using the rewriting rules similarly as in~\cite{MMS07}. However, our experiments show that, using the value index, we can also efficiently process queries with value predicates.

In our prototype, we have implemented three approaches to process a~TPQ: a~holistic join \algo{GTPStack} \cite{Ba+13}, a~cost-based holistic join \algo{CostTwigJoin} \cite{bavca2015cost}, and fully-pipelined query plans of binary structural joins (FP-BJ) \cite{lukavs2017demythization}. All of the three approaches require the detection of a~TPQ. \algo{GTPStack} represents a~state-of-the-art holistic join and it utilizes the partition index or the value index to retrieve lists of XML nodes specified by query nodes or value predicates, respectively, \algo{CostTwigJoin} extends \algo{GTPStack} to combine the partition and document index, which is efficient especially for highly selective queries. The FP-BJ approach improves the binary join query processing; it uses the partition index and the value index and it is especially advantageous for XQueries with a~low number of output query nodes.


\section{Experiments}
\label{sec:experiments}

We picked 5 native XML DBMSs with the highest ranking according to \url{http://db-engines.com} (September 2018). We added MonetDB, which we consider as one of the most efficient native XML DBMSs despite the fact that it is no longer supported. We also included two relational DBMSs supporting XML querying. All of these XML DBMSs are summarized in Table~\ref{tab:tested_dbmss}. The table also includes abbreviations we use in the following text. For the license purposes, we cannot provide the names of the both relational DBMSs (CR1, CR2) and one of the XML DBMSs (CX). We employed five XML data collections used by many approaches~\cite{MMS07, ch03, GBH10, Ba+13}. XMark is a well-known XML benchmarking synthetic collection whose size can be controlled using a~factor $f$. We worked with two factors $f = 1$ and $f = 10$. SwissProt is a~real database of protein sequences, TreeBank contains partially encrypted English sentences and DBLP is a~real bibliographical database. These collections and their statistics are summarized in Table~\ref{tab:collections}. Unfortunately, the DBMSs CX and E-DB were not able to store any of the tested XML collections. Therefore, these two DBMSs were omitted from the following experiments. For the other DBMSs, we created structural and value indexes, if it was possible. We used the Intel Xeon E5-2690@2.9GHz processor, 384\,GB RAM, and the Microsoft Windows Server 2016 Datacenter operating system as the testbed.

\begin{table}[htb]
	\scriptsize
	\begin{tabular}{|p{8cm}|}
		\hline
		\textbf{Oracle Berkley DB 6.1.4} (B-DB) \newline
		\url{www.oracle.com/­technetwork/­database/­database-technologies/­berkeleydb/­overview/­index.html} \\
		\hline
		\textbf{Virtuoso 7.1} (VRT) \newline
		\url{virtuoso.openlinksw.com} \\
		\hline
		\textbf{eXist-db 4.3.1} (E-DB) \newline
		\url{exist-db.org} \\
		\hline
		\textbf{BaseX 9.0.2} (BX) \newline
		\url{basex.org} \\
		\hline
		\textbf{MonetDB XQuery 4} (M-DB) \newline
		\url{www.monetdb.org/XQuery} \\
		\hline
		\textbf{Commercial XML DBMS} (CX) \\
		\hline
		\textbf{Commercial relational DBMS 1 and 2} (CR1, CR2)
		\\
		\hline
	\end{tabular}
	\caption{DBMSs included in experiments}
	\label{tab:tested_dbmss}	
\end{table}


\begin{table}[htb]
	\scriptsize
	\centering
	\begin{tabular}{|l|rrr|}
		\hline
		\textbf{Collection} & \textbf{Size (MB)} & \textbf{XML nodes} & \textbf{Max. depth} \\
		\hline
		\hline
		XMark (f=1) & 111 & 2,048,193 & 14 \\
		XMark (f=10) & 1,137 & 20,532,805 & 14 \\
		SwissProt & 109 & 5,166,890 & 7 \\
		TreeBank & 82 & 2,437,667 & 38 \\
		DBLP & 127 & 3,736,406 & 6 \\		
		\hline
	\end{tabular}
	\caption{Statistics of data collections}
	\label{tab:collections}
\end{table}

For each collection in Table~\ref{tab:collections} we prepared 5 structural XQueries and 5 XQueries with value predicates; all these queries are listed in Appendix~\ref{sec:structural_queries} and Appendix~\ref{sec:content_queries}, respectively.
We measured the query processing time without the result materialization (i.e., without retrieving the whole subtrees of data nodes searched by the queries). Therefore, we wrapped the queries with the XQuery function \texttt{count()}. The queries were run 5 times on each database; we considered the arithmetic means of the processing times without the best and the worst run. The processing times for all the queries, collections and DBMSs can be found in Table~\ref{tab:processing_times}. The best processing times are in bold.

\begin{figure}[b]
	\centering
	\small
	\begin{tabular}{cc}
		\hspace{-0.7cm} \raisebox{3.3cm}{$(s) \uparrow$} \hspace{-0.5cm} &
		\includegraphics[width=8cm]{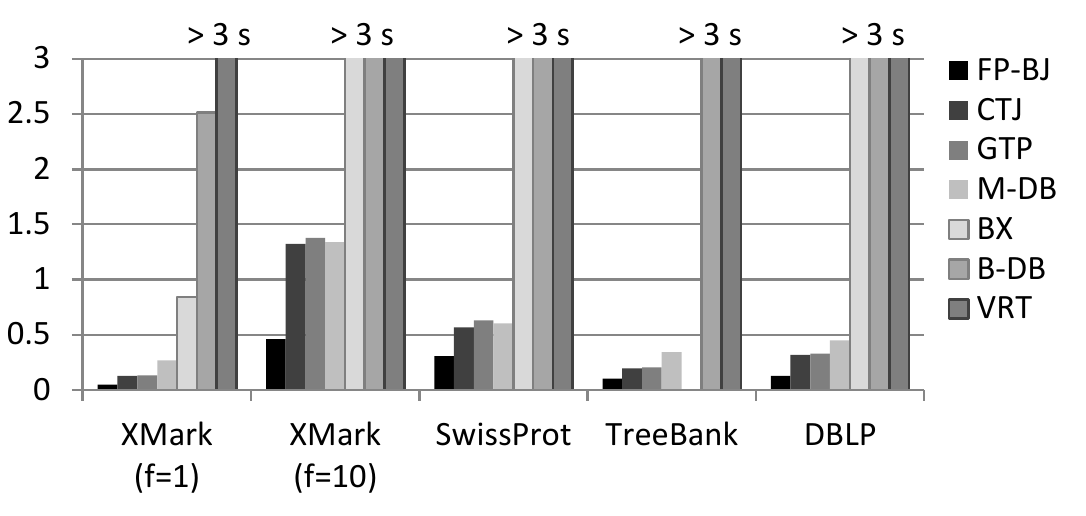} 
		\hspace{-0.5cm}
	\end{tabular}
	\caption{Processing times of structural queries}
	\label{fig:structural_times}
\end{figure}

Let us first focus on the structural queries whose results are summarized in Figure~\ref{fig:structural_times}. The FP-BJ, CTJ (\algo{CostTwigJoin}), and GTP (\algo{GTPStack}) represent the approaches implemented in our prototype. We omitted the relational databases CR1 and CR2 in the figure, since the most of the queries did not finished until 5 minutes. For the same reason, we also omitted the results of BX for the TreeBank collection.

We can observe, that our FB-BJ approach gives the best overall performance on all collections. It processed 23 (out of 25) queries with the best processing time compared to the other DBMSs in Table~\ref{tab:tested_dbmss}, and 17 queries with the best time (including CTJ and GTP in the comparison). FP-BJ performed from $1.36 \times$ to $15.49 \times$ better than the best of the other DBMSs in Table~\ref{tab:tested_dbmss} which was usually M-DB. 7~queries were processed most efficiently by CTJ and 1 query by M-DB.

\begin{figure}[htb]
	\centering
	\small
	\begin{tabular}{cc}
		\hspace{-0.7cm} \raisebox{3.7cm}{$(s) \uparrow$} \hspace{-0.5cm} &
		\includegraphics[width=8cm]{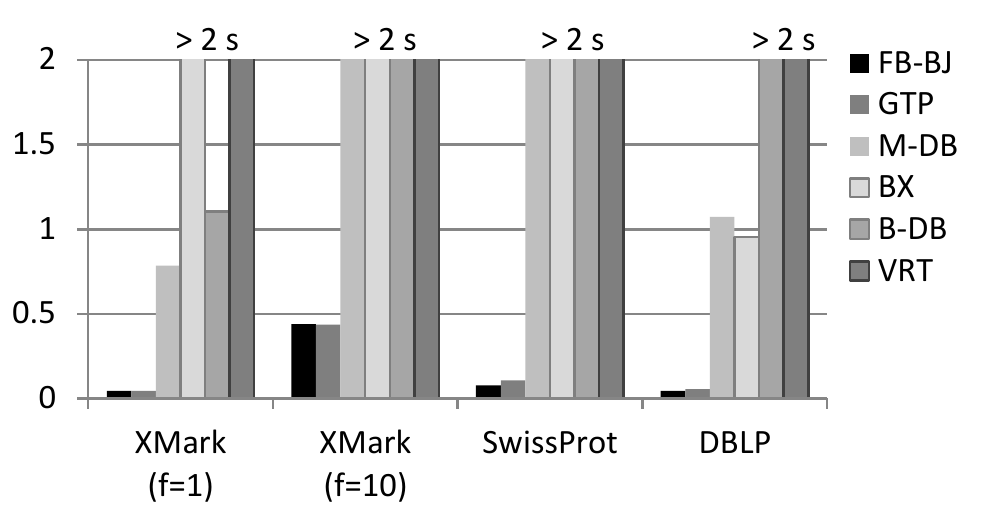} 
		\hspace{-0.5cm}
	\end{tabular}
	\caption{Processing times of queries with value predicates}
	\label{fig:content_times}
\end{figure}

For queries with value predicates, we omitted the TreeBank collection because of its encrypted content for which it is not reasonable to perform such queries. We also do not include results of CTJ, since it requires statistics of distribution of values which are currently not supported in our prototype. Summarized results for queries with value predicates are in Figure~\ref{fig:content_times}.

Similarly as in the previous experiment, we can see that FP-BJ gives the best overall performance on all collections. It processed 18 (out of 20) queries with the best processing time compared to the other DBMSs in Table~\ref{tab:tested_dbmss}, and 15 queries with the best time (including GTP in the comparison). FP-BJ performed from $2.60 \times$ to $130.43 \times$ better than the best of the other DBMSs in Table~\ref{tab:tested_dbmss}. 2 queries were processed most efficiently by BX, but M-DB gave the best overall performance out of the other tested DBMSs again.


Finally, we tried to find out whether the other DBMSs utilize any detection of a~TPQ. We measured query processing times of two semantically equivalent queries (these two times should be nearly the same, if the same TPQ is detected) and we examined plans of these queries. We realized that any of the DBMSs in Table~\ref{tab:tested_dbmss} do not utilize any detection of a~TPQ.


\begin{table*}
	\centering
	\scriptsize
	{
		\setlength{\tabcolsep}{0.4em}
		\renewcommand{\arraystretch}{0.8}
		\begin{tabular}{|c|rrr|rrrrrr||rr|rrrrrr|}
			\hline
			&
			\multicolumn{9}{c||}{\textbf{Structural queries}}
			&
			\multicolumn{8}{c|}{\textbf{Queries with value predicates}} \\
			\hline
			& GTP & CTJ & FP-BJ & B-DB & VRT & BX & M-DB & CR1 & CR2 & GTP & FP-BJ & B-DB & VRT & BX & M-DB & CR1 & CR2 \\
			\hline
			XM1&0.010&0.011&\textbf{0.002}&0.265&4.422&0.112&0.030&63.655&DNF&\textbf{0.004}&0.008&0.474&4.213&0.347&0.145&93.868&DNF\\
			XM2&0.041&0.041&\textbf{0.016}&0.870&4.979&0.358&0.094&DNF&DNF&\textbf{0.023}&\textbf{0.023}&0.193&4.156&0.981&0.111&9.710&DNF\\
			XM3&0.007&\textbf{0.003}&0.010&0.078&4.318&0.013&0.036&0.054&0.031&\textbf{0.000}&\textbf{0.000}&0.021&4.094&0.004&0.139&2.140&DNF\\
			XM4&0.041&0.041&\textbf{0.016}&0.740&4.922&0.137&0.062&DNF&DNF&\textbf{0.010}&\textbf{0.010}&0.084&3.984&2.508&0.134&3.510&DNF\\
			XM5&0.034&0.033&\textbf{0.008}&0.563&4.630&0.223&0.045&223.361&DNF&0.006&\textbf{0.002}&0.333&3.990&0.168&0.256&DNF&DNF\\
			\hline
			XM1&0.108&0.105&\textbf{0.025}&2.427&58.573&0.972&0.106&DNF&DNF&\textbf{0.031}&0.068&9.120&50.057&3.479&0.665&DNF&DNF\\
			XM2&0.423&0.423&\textbf{0.141}&8.450&66.136&3.079&0.621&DNF&DNF&0.236&\textbf{0.211}&1.677&51.078&3.377&0.548&DNF&DNF\\
			XM3&0.071&\textbf{0.032}&0.078&0.250&59.500&0.087&0.063&DNF&0.250&0.006&\textbf{0.000}&0.021&50.141&0.005&0.528&16.494&DNF\\
			XM4&0.436&0.426&\textbf{0.141}&8.146&71.620&1.328&0.355&DNF&DNF&0.121&\textbf{0.117}&0.370&51.349&9.451&0.516&253.375&DNF\\
			XM5&0.338&0.337&\textbf{0.074}&5.468&66.354&1.857&0.195&DNF&DNF&\textbf{0.041}&\textbf{0.041}&5.292&51.588&1.547&1.073&DNF&DNF\\
			\hline
			SP1&0.019&\textbf{0.006}&0.016&0.042&9.828&1.030&0.095&DNF&10.083&\textbf{0.000}&\textbf{0.000}&0.021&9.828&0.002&0.166&6.197&DNF\\
			SP2&0.210&0.199&\textbf{0.100}&3.943&15.156&1.708&0.170&DNF&DNF&\textbf{0.002}&\textbf{0.002}&3.500&10.224&1.103&0.254&0.434&DNF\\
			SP3&0.042&0.042&\textbf{0.020}&1.656&10.375&0.675&0.136&DNF&DNF&0.063&\textbf{0.055}&4.933&10.135&1.280&DNF&DNF&DNF\\
			SP4&0.021&\textbf{0.012}&0.016&0.062&9.719&0.574&0.052&19.716&0.297&\textbf{0.000}&\textbf{0.000}&1.026&9.776&1.521&0.573&1.303&DNF\\
			SP5&0.338&0.306&0.156&9.073&16.797&2.821&\textbf{0.150}&DNF&DNF&0.041&\textbf{0.020}&0.313&9.411&0.276&0.963&54.766&DNF\\
			\hline
			TB1&0.012&\textbf{0.002}&0.012&0.469&6.073&DNF&0.048&DNF&130.187&&&&&&&&\\
			TB2&0.016&\textbf{0.014}&0.016&78.651&6.375&DNF&0.091&DNF&DNF&&&&&&&&\\
			TB3&\textbf{0.012}&\textbf{0.012}&0.016&0.766&5.255&DNF&0.059&DNF&DNF&&&&&&&&\\
			TB4&0.068&0.068&\textbf{0.018}&0.740&5.359&DNF&0.058&DNF&0.109&&&&&&&&\\
			TB5&0.099&0.100&\textbf{0.043}&6.136&6.672&DNF&0.088&DNF&DNF&&&&&&&&\\
			\hline
			DB1&0.066&0.064&\textbf{0.031}&3.693&10.089&1.435&0.084&DNF&DNF&0.012&0.006&8.338&9.214&\textbf{0.002}&0.154&61.869&DNF\\
			DB2&0.011&0.011&\textbf{0.010}&0.047&8.672&0.511&0.035&14.162&0.141&\textbf{0.000}&\textbf{0.000}&0.922&8.870&0.028&0.208&26.061&DNF\\
			DB3&0.022&0.021&\textbf{0.016}&1.875&9.031&0.881&0.043&DNF&DNF&0.016&\textbf{0.008}&6.000&9.011&0.804&0.269&28.120&DNF\\
			DB4&0.188&0.184&\textbf{0.059}&5.031&10.036&1.741&0.227&DNF&DNF&0.010&0.012&3.521&9.078&\textbf{0.005}&0.208&4.603&DNF\\
			DB5&0.040&0.039&\textbf{0.014}&0.411&9.015&1.081&0.061&DNF&7.630&\textbf{0.016}&0.018&1.531&9.302&0.112&0.234&176.474&DNF\\
			\hline
		\end{tabular}
	}
	\vspace{-1em}
	\flushleft \emph{DNF -- did not finished until 5 minutes}
	
	\caption{Processing time of structural queries and queries with value predicates in seconds}
	\label{tab:processing_times}
\end{table*}

\section{Conclusions}
\label{sec:conclusions}

Our experiments clearly show that our prototype, RadegastXDB, performs better compared to the current XML DBMSs for both structural queries and queries with value predicates. This is caused by utilizing state-of-the-art XML indexing techniques and TPQ processing algorithms. The experiments also show that the current XML DBMSs ignore the most of the advances in the XML query processing introduced during the past years. For example, holistic twig joins have been introduced in 2002 \cite{BSK02}, and since then, they have been considered as a state-of-the-art XML query processing technique. Now, in 2019, they are still not integrated in any of the current XML DBMSs. Although, there is still a~plenty of future work on our prototype including, e.g., the complete support of XQuery or the transactional processing, we believe that it already gives valuable results.

\bibliographystyle{abbrv}
\bibliography{references}

\clearpage
\appendix

\twocolumn[
\section{Structural Queries}
\label{sec:structural_queries}

\scriptsize
\centering
\begin{tabular}{p{7.5cm}p{7.5cm}}

\textbf{XMark} \newline
\noindent \begin{tabular}{|c|m{6cm}|}
	\hline
	\textbf{XM1} & \texttt{\textbf{for} \$a \textbf{in} //open\_auctions/open\_auction \linebreak \textbf{for} \$b \textbf{in} \$a/interval \linebreak \textbf{where} \$b//start \linebreak \textbf{return} (\$a, \$b)}
	\linebreak (result size: 24,000 / 240,000)
	\\
	\hline
	\textbf{XM2} & \texttt{\textbf{for} \$a \textbf{in} //regions//item \linebreak \textbf{for} \$b \textbf{in} \$a//text \linebreak \textbf{for} \$c \textbf{in} \$b/keyword \linebreak \textbf{where} \$a/mailbox/mail[./to \textbf{and} ./from] \linebreak \textbf{return} (\$a, \$b, \$c)}
	\linebreak (result size: 80,925 / 812,838)
	\\
	\hline
	\textbf{XM3} & \texttt{\textbf{for} \$a \textbf{in} //regions/samerica/item \linebreak \textbf{where} \$a/mailbox/mail/to \textbf{and} \$a/incategory/@category \linebreak \textbf{return} \$a}
	\linebreak (result size: 582 / 6073)
	\\
	\hline
	\textbf{XM4} & \texttt{\textbf{for} \$a \textbf{in} //people/person/profile \linebreak \textbf{for} \$b \textbf{in} \$a/interest \linebreak \textbf{for} \$c \textbf{in} \$b/@category \linebreak \textbf{where} \$a//business \linebreak \textbf{return} (\$a, \$b, \$c)}
	\linebreak (result size: 113,067 / 1,141,008)
	\\
	\hline
	\textbf{XM5} & \texttt{\textbf{for} \$a \textbf{in} //regions//item \linebreak \textbf{for} \$b \textbf{in} \$a/location \linebreak \textbf{where} \$a/description/text \textbf{and} \$a/shipping \linebreak \textbf{return} (\$a, \$b)}
	\linebreak (result size: 30,904 / 308,428)
	\\
	\hline
\end{tabular}
\emph{The two result sizes for XM queries correspond to the two factors of the XMark collection ($f=1, f=10$).}	

&

\textbf{TreeBank} \newline
\noindent \begin{tabular}{|c|m{6cm}|}
	\hline
	\textbf{TB1} & \texttt{\textbf{for} \$a \textbf{in} //EMPTY \linebreak \textbf{for} \$b \textbf{in} \$a//X \linebreak \textbf{where} \$a/\_PERIOD\_ \textbf{and} \$b//\_COLON\_ \textbf{and} \$b//NP//NNPS \linebreak \textbf{return} (\$a, \$b)}
	\linebreak (result size: 20)
	\\						
	\hline
	\textbf{TB2} & \texttt{\textbf{for} \$a \textbf{in} //EMPTY \linebreak \textbf{for} \$b \textbf{in} \$a/S \linebreak \textbf{for} \$c \textbf{in} \$b//VP \linebreak \textbf{where} \$c//VB \textbf{and} \$c/PP[.//VBG \textbf{and} ./TO] \linebreak \textbf{return} (\$a, \$b, \$c)}
	\linebreak (result size: 318)
	\\						
	\hline
	\textbf{TB3} & \texttt{\textbf{for} \$a \textbf{in} //EMPTY \linebreak \textbf{for} \$b \textbf{in} \$a//S/SINV/VP//VBZ \linebreak \textbf{where} \$a/\_PERIOD\_ \linebreak \textbf{return} (\$a, \$b)}
	\linebreak (result size: 1,864)
	\\						
	\hline
	\textbf{TB4} & \texttt{\textbf{for} \$a \textbf{in} //EMPTY//VP/NP//NNS \linebreak \textbf{return} \$a}
	\linebreak (result size: 33,971)
	\\						
	\hline
	\textbf{TB5} & \texttt{\textbf{for} \$a \textbf{in} //EMPTY//NP \linebreak \textbf{for} \$b \textbf{in} \$a/JJ \linebreak \textbf{for} \$c \textbf{in} \$a//NN \linebreak \textbf{return} (\$a, \$b, \$c)}
	\linebreak (result size: 159,591)
	\\						
	\hline
\end{tabular}

\\

\textbf{SwissProt} \newline
\noindent \begin{tabular}{|c|m{6cm}|}
	\hline
	\textbf{SP1} & \texttt{\textbf{for} \$a \textbf{in} //Entry \linebreak \textbf{where} \$a/@mtype \textbf{and} \$a//SUBTILIST[./@sec\_id \textbf{and} ./@prim\_id] \linebreak \textbf{return} \$a}
	\linebreak (result size: 859)
	\\						
	\hline
	\textbf{SP2} & \texttt{\textbf{for} \$a \textbf{in} //Entry \linebreak \textbf{for} \$b \textbf{in} \$a/Ref//Comment \linebreak \textbf{for} \$c \textbf{in} \$a//Descr \linebreak \textbf{return} (\$a, \$b, \$c)}
	\linebreak (result size: 1,240,896)
	\\						
	\hline
	\textbf{SP3} & \texttt{\textbf{for} \$a \textbf{in} //Entry[./@id] \linebreak \textbf{for} \$b \textbf{in} \$a/HSSP[./@prim\_id \textbf{and} ./@sec\_id] \linebreak \textbf{return} (\$a, \$b)}
	\linebreak (result size: 34,134)
	\\						
	\hline
	\textbf{SP4} & \texttt{\textbf{for} \$a \textbf{in} //Entry \linebreak \textbf{where} \$a/TIGR/@prim\_id \textbf{and} \$a/EMBL/@sec\_id \linebreak \textbf{return} \$a}
	\linebreak (result size: 1,751)
	\\						
	\hline
	\textbf{SP5} & \texttt{\textbf{for} \$a \textbf{in} //Entry//Ref \linebreak \textbf{for} \$b \textbf{in} \$a/Author \linebreak \textbf{for} \$c \textbf{in} \$a/MedlineID \linebreak \textbf{return} (\$a, \$b, \$c)}
	\linebreak (result size: 1,465,377)
	\\						
	\hline
\end{tabular}				

&

\textbf{DBLP} \newline
\noindent \begin{tabular}{|c|m{6cm}|}
	\hline
	\textbf{DB1} & \texttt{\textbf{for} \$a \textbf{in} //inproceedings[.//pages] \linebreak \textbf{for} \$b \textbf{in} \$a//cite \linebreak \textbf{return} (\$a, \$b)}
	\linebreak (result size: 541,644)
	\\						
	\hline
	\textbf{DB2} & \texttt{\textbf{for} \$a \textbf{in} //article[./month]//year \linebreak \textbf{return} \$a}
	\linebreak (result size: 2,474)
	\\						
	\hline
	\textbf{DB3} & \texttt{\textbf{for} \$a \textbf{in} //inproceedings \linebreak \textbf{for} \$b \textbf{in} \$a//cdrom \linebreak \textbf{for} \$c \textbf{in} \$a/url \linebreak \textbf{return} (\$a, \$b, \$c)}
	\linebreak (result size: 29,766)
	\\						
	\hline
	\textbf{DB4} & \texttt{\textbf{for} \$a \textbf{in} //inproceedings[./year \textbf{and} ./author] \linebreak \textbf{for} \$b \textbf{in} \$a//crossref \linebreak \textbf{return} (\$a, \$b)}
	\linebreak (result size: 222,594)
	\\						
	\hline
	\textbf{DB5} & \texttt{\textbf{for} \$a \textbf{in} //inproceedings[.//crossref]//ee \linebreak \textbf{return} \$a}
	\linebreak (result size: 47,624)
	\\						
	\hline
\end{tabular}
\\
\end{tabular}]

\clearpage

\twocolumn[
\section{Queries with Value Predicates}
\label{sec:content_queries}

\centering
\scriptsize
\begin{tabular}{p{7.5cm}p{7.5cm}}

\textbf{XMark} \newline
\noindent \begin{tabular}{|c|m{6cm}|}
	\hline
	\textbf{XM1} & \texttt{\textbf{for} \$a \textbf{in} //item \linebreak
		\textbf{for} \$b \textbf{in} \$a/mailbox/mail \linebreak
		\textbf{let} \$c := \$b/from \linebreak
		\textbf{let} \$d := \$b/to \linebreak
		\textbf{where} \$a/location = "United States" \textbf{and} \$a/quantity > 2 \linebreak
		\textbf{return} (\$a, \$b, \$c)}
	\linebreak (result size: 174 / 2,433)
	\\
	\hline
	\textbf{XM2} & \texttt{\textbf{for} \$a \textbf{in} //item \linebreak
		\textbf{where} \$a/name >= "a" \textbf{and} \$a/name < "d" \linebreak
		\textbf{return} \$a}
	\linebreak (result size: 4,454 / 44,849)
	\\
	\hline
	\textbf{XM3} & \texttt{//open\_auction[privacy = "No" \textbf{and} itemref/@item = "item10"]//keyword} 
	\linebreak (result size: 1 / 0)
	\\
	\hline
	\textbf{XM4} & \texttt{\textbf{for} \$a \textbf{in} //closed\_auction[.//happiness = 5] \linebreak
		\textbf{where} \$a/quantity < 3 \linebreak
		\textbf{return} \$a}
	\linebreak (result size: 979 / 9,717)
	\\
	\hline
	\textbf{XM5} & \texttt{//person[address/country="United States" \textbf{and} profile[gender="female" \textbf{and} age >= 20 \textbf{and} age <= 25]]} 
	\linebreak (result size: 92 / 860)
	\\
	\hline
\end{tabular}
\emph{The two result sizes for XM queries correspond to the two factors of the XMark collection ($f=1, f=10$).}
&
\textbf{DBLP} \newline
\noindent \begin{tabular}{|c|m{6cm}|}
	\hline
	\textbf{DB1} & \texttt{//inproceedings[./author = "Joel Wein"]/title}
	\linebreak (result size: 13)
	\\						
	\hline
	\textbf{DB2} & \texttt{//article[./journal="IEEE Transactions on Computers" \textbf{and} number=2]}
	\linebreak (result size: 411)
	\\						
	\hline
	\textbf{DB3} & \texttt{//inproceedings[./year=1998]/booktitle}
	\linebreak (result size: 16,235)
	\\						
	\hline
	\textbf{DB4} & \texttt{\textbf{for} \$a \textbf{in} //article \linebreak
		\textbf{for} \$b \textbf{in} \$a/journal \linebreak
		\textbf{where} \$a/author = "Jennifer Widom" \textbf{and} \$a//url \linebreak
		\textbf{return} \$b}
	\linebreak (result size: 38)
	\\						
	\hline
	\textbf{DB5} & \texttt{\textbf{for} \$a \textbf{in} //article \linebreak
		\textbf{for} \$b \textbf{in} \$a/author \linebreak
		\textbf{where} \$a/number = 3 \textbf{and} \$a/volume = 6 \linebreak
		\textbf{return} \$b}
	\linebreak (result size: 1,240)
	\\						
	\hline
\end{tabular}
\\

\vspace{1em}
\textbf{SwissProt} \newline
\noindent \begin{tabular}{|c|m{6cm}|}
	\hline
	\textbf{SP1} & \texttt{//Entry[.//Author = "Rehbein M" \textbf{and} Org = "Muridae"]/AC}
	\linebreak (result size: 4)
	\\						
	\hline
	\textbf{SP2} & \texttt{\textbf{for} \$a \textbf{in} //Entry \linebreak
		\textbf{for} \$b \textbf{in} \$a//INTERPRO/@prim\_id \linebreak
		\textbf{for} \$c \textbf{in} \$a//PFAM/@prim\_id \linebreak
		\textbf{where} \$b = "IPR000569" \textbf{and} \$c = "PF00632" \linebreak
		\textbf{return} \$a}
	\linebreak (result size: 6)
	\\						
	\hline
	\textbf{SP3} & \texttt{\textbf{for} \$a \textbf{in} //Entry \linebreak
		\textbf{for} \$b \textbf{in} \$a//Mod[@Rel > 10 \textbf{and} @Rel < 15 \textbf{and} @type = "Created"] \linebreak
		\textbf{for} \$c \textbf{in} \$a/Gene \linebreak
		\textbf{return} (\$a, \$c)}
	\linebreak (result size: 5,300)
	\\						
	\hline
	\textbf{SP4} & \texttt{\textbf{for} \$a \textbf{in} //Entry \linebreak
		\textbf{for} \$b \textbf{in} \$a/Ref \linebreak
		\textbf{where} \$a/Org = "Eukaryota" \textbf{and} \$b/Author = "Piotrowski M" \linebreak
		\textbf{return} \$b}
	\linebreak (result size: 5)
	\\						
	\hline
	\textbf{SP5} & \texttt{//Entry[Species = "Homo sapiens (Human)" \textbf{and} Ref[@num <= 1]]//Descr}
	\linebreak (result size: 61,946)
	\\						
	\hline
\end{tabular} \\

\end{tabular} 
]

\end{document}